\documentclass[12pt]{article}
\usepackage[dvips]{graphicx}
\usepackage{amsmath}
\usepackage{array}
\usepackage{latexsym,setspace}
\usepackage{subfigure}
\usepackage{fancyhdr}
\usepackage{bm}

\begin{document}
\begin{center}
\LARGE{\textbf{Self-Interacting Dark Matter in the \\Solar System}}\\

\vspace{15mm}
\large{
\textbf{Avijit K. Ganguly$^1$, Pankaj Jain$^2$, Subhayan Mandal$^2$ and Sarah Stokes$^3$}}\\

\vspace{10mm}
{$^1$ Haldia Institute of Technology, Haldia, India}\\
{$^2$Department of Physics\\
Indian Institute of Technology \\
Kanpur 208016, INDIA}\\
{$^3$Department of Physics\\
California Institute of Technology \\
Pasadena, CA, USA}\\

\bigskip
\today
\end{center}

\vspace{5mm}
\noindent
\textbf{Abstract:} 
Weakly coupled, almost massless, spin 0 particles have been predicted
by many extensions of the standard model of particle physics.
Recently, the PVLAS group observed a rotation of polarization
of electromagnetic waves in vacuum in the presence of transverse
magnetic field.  This phenomenon is best explained by the existence of
a weakly coupled light pseudoscalar particle.
However, the coupling required by this experiment is much larger than the
conventional astrophysical limits. 
Here we consider a hypothetical self-interacting
pseudoscalar particle which couples weakly with visible matter. 
 Assuming that these pseudoscalars pervade 
the galaxy, we show that the solar limits 
on the pseudoscalar-photon coupling can be evaded.

\section{Introduction}
In a recent paper Jain and Mandal \cite{Jain06} 
proposed a mechanism to evade the 
astrophysical bounds on light pseudoscalars. The authors argued that there
are no experimental or astrophysical bounds on the pseudoscalar self-couplings,
which can, therefore, be treated as a free parameter. For sufficiently large, 
but perturbative, self-coupling the mean free path of pseudoscalars can be
very small inside the sun, provided we have sufficiently high density of these
particles. The authors argued that the pseudoscalars produced inside the 
sun will accumulate due to self-couplings. The mechanism required the
fragmentation of the pseudoscalars $\phi$ through the process 
$\phi(k_1)+\phi(k_2)
\rightarrow \phi(k_3)+\phi(k_4) + \phi(k_5)+\phi(k_6)$. 
Some mechanism for energy loss as the particles propagate
through the sun was also required in order that the pseudoscalar particles
are at sufficiently low temperature. Using this mechanism the authors argued
the results of the PVLAS experiment \cite{PVLAS}
can be consistent with the astrophysical 
bounds \cite{astrobounds} and with the results of the 
CAST experiment \cite{CAST}. The PVLAS collaboration finds a 
rotation of polarization
of light in vacuum in the presence of a transverse magnetic field. If
we interpret this rotation in terms of the coupling of a
light pseudoscalar particle to photons
we find that the allowed range of parameters are in conflict with the 
astrophysical bounds \cite{astrobounds}, although there is no conflict with
any laboratory bounds \cite{Cameron}. The PVLAS result has motivated
considerable theoretical work [6-15] as well as proposals for new experiments
\cite{Jaeckel,Koetz} and observations \cite{Fairbairn}. 
Furthermore new contraints have been imposed on axion monopole-dipole coupling
\cite{Baessler}
assuming the mass range observed in the PVLAS experiment.

Self-interacting dark matter has earlier been considered by many authors.
It was proposed in Ref. \cite{Carlson} due to its interesting consequences
for cosmology. Constraints on such a dark matter candidate and 
its implications for
cosmology have been further studied \cite{self1}. The dark matter studied
in the present paper may not give a significant contribution to the energy
density of the universe due to its very small mass. 
Significant contribution may arise only if the pseudoscalar field undergoes
coherent oscillations around the minimum of the potential such as those
expected for axion like fields \cite{coherent}. This oscillatory
field behaves just as non-relativistic dark matter. Some evidence for
self-interacting dark matter is also found in considering the galactic 
dark matter distribution \cite{self2}. The standard cold dark matter 
scenario leads to cuspy dark matter galactic halos, which are not in
agreement with observations \cite{Swaters}. 
The self interacting dark matter solves this
problem and has been studied by many authors \cite{self3}. 
Alternatively in Ref. \cite{Verde} it is claimed that the dark matter profile
is in good agreement with cold dark matter predictions.
It is interesting to determine whether the dark matter we consider can
be the dominant component of the galactic dark matter. However we
postpone this question for future research and for now simply assume that the
pseudoscalars we consider have densities negligible compared to the
galactic dark matter density.

In the present paper we propose a simple mechanism for evading the 
astrophysical bounds. Our basic assumptions are
\begin{itemize}
\item[1.] The pseudoscalars have a perturbative, but relatively large,
self-coupling. 
\item[2.] The pseudoscalars form some component of the galactic dark matter.
\end{itemize}
The second assumption implies that pseudoscalars at nearly zero temperature
are present throughout the galaxy. This assumption is quite reasonable since
pseudoscalars will be produced in the early universe and will be present today
in the form of dark matter. 
Stars are formed in regions of dense molecular clouds where the average
density of matter is much higher then the mean galactic density. 
It is very likely that the density of dark matter is also higher
in these regions
due to the large gravitational attraction. As the cloud of 
dust and gas collapses to form a star the pseudoscalar cloud will also 
collapse. This will lead to a high density of pseudoscalars inside the star.
The mean free path of pseudoscalars produced inside the star may then
be very small due to the their scattering on background pseudoscalars
by the reaction $\phi(k_1)+\phi(k_2)
\rightarrow \phi(k_3)+\phi(k_4)$. Hence this mechanism can trap pseudoscalars
inside the star and and considerably limit the radiative transfer
\cite{Raffelt} that can
occur through pseudoscalars.
In the present paper we investigate this mechanism in order to 
determine whether this mechanism can indeed evade the astrophysical 
bounds. We will focus on the bounds imposed by considering
the pseudoscalar production inside the core of the sun. Our mechanism also
applies to bounds due to red giants and supernovae. However we postpone this
discussion to future research. 

\section{Evading the Solar Bounds}
The effective Lagrangian for the pseudoscalars can be written as
\begin{equation}
{\cal L} = -{1\over 4} F_{\mu\nu}F^{\mu\nu}+{1\over 2}\partial_\mu\phi
\partial^\mu \phi - {1\over 2}m^2 \phi^2 +  
{1\over 4 M_\phi} \phi F_{\mu\nu}\tilde
F^{\mu\nu} - {\lambda\over 4! } \phi^4
\label{eq:Lphi_I}
\end{equation}
where we have assumed a self-coupling term besides their interaction with
photons. Assuming that the results of PVLAS experiment can be explained 
in terms of pseudoscalar-photon mixing, the parameter $M_\phi$ and the 
pseudoscalar mass $m_\phi$, respectively, lie in the range
$1\times 10^5\ {\rm GeV} \le M_\phi \le 6\times 10^5\ {\rm GeV}$ and
        $0.7\ {\rm meV} \le m_\phi \le 2\ {\rm meV}$ \cite{PVLAS}.
We assume that the self-coupling $\alpha_\lambda = \lambda^2 /4\pi 
\le 0.1$, so that it can be treated perturbatively.

The density of dark matter in the galaxy may be parametrized as \cite{ostlie},
\begin{equation}
\rho(r) =\frac{C_o}{a^2+r^2}
\label{eqn:gdensity}
\end{equation}
We may assume that the pseudoscalars form a fraction $\xi$ of the
galactic dark matter,
where $C_o = 4.6\times10^8 M_{\odot} kpc^{-1}$ and $a=2.8$ kpc.
Using the mass of the pseudoscalar to be about
$10^{-3}$ eV and the distance of earth from the galactic center 
$r = 8$ kpc, we estimate the number density
$n= 2.4\xi  10^{11} cm^{-3}$
We assume $\xi$ to be less than 0.1 so that pseudoscalars contribute 
negligibly to the galactic dark matter density.
This parametrization of the density profile is sufficiently reliable for
our purpose. It correctly
models the expected $1/r^2$ at large distances and becomes constant at small
distances. The simulated 
dark matter distributions for several different models such
as the standard cold dark matter and cold dark matter with self
interaction are shown in Ref. \cite{self2}.
 The resulting density obtained for large $r$ is
in good agreement with that obtained with the parameterization given in 
eq. \ref{eqn:gdensity} for these models. 
 Hence we find that a wide range of reasonable models
of galactic dark matter produce a density in good agreement with what is
obtained from eq. \ref{eqn:gdensity} at the position of the earth inside the milky way.

The pseudoscalar density may be higher at the site of star
formation due to the higher concentration of matter in this region. 
We set the pseudoscalar number density in region which
lead to the formation of sun to be a factor
$\eta$ times its galactic density. This region, which contained the
primordial cloud of gas and dust whose collapse lead to the formation
of the solar system, is expected to be of size larger than 
100 - 1000 AU. As the star forms the pseudoscalars in this region may 
also collapse onto the star. 
The precise pseudoscalar density profile will be determined later by requiring
that the solar system, including the pseudoscalar gas, are in steady state. 
Here we next make an order of magnitude estimate of the pseudoscalar number
density inside the sun, assuming that almost all the pseudoscalars within
a radius $R_S \approx 1000$ AU collapse onto the sun. The resulting number
density of pseudoscalars inside the sun is found to be
\begin{equation}
n_{\phi\odot} \approx 2.4 \xi\eta 10^{26} \left[{R_S\over 1000 {\rm AU}}\right]^3
{\rm cm}^{-3}
\label{eqn:nden}
\end{equation}

As the star collapses the temperature in the core increases, eventually
leading to nuclear reactions. The photons in the core then start producing
pseudoscalars through the Primakoff process \cite{Jain06}. The temperature
of the pseudoscalars inside the sun also increases, partially due to the
collapse of pseudoscalars and partially due to production of high energy
pseudoscalars by the Primakoff process. We can compute the kinetic energy
of pseudoscalars as they reach the solar radius from nearly infinite 
distance. Assuming that their initial kinetic energy is almost zero we
find that the final kinetic energy is $3.4\times 10^{-21}$ ergs which is
equivalent to a temperature of $2.5\times 10^{-5}$ K. Hence this predicts a
relatively cool pseudoscalar gas. This result may be 
significantly modified due to conversion of photons into pseudoscalars inside
the sun. As already mentioned the final density profile is obtained later
 assuming
steady state of the solar system and does not rely on these rough estimates.

The typical energy of a pseudoscalar
produced inside the core of the sun, assuming present day conditions,
is of order 1 KeV. This is much larger than the kinetic energy acquired
by pseudoscalars due to conversion of potential energy. We expect that
the pseudoscalars will eventually reach a steady state, where the outward
pressure balances the gravitational attraction. Furthermore there
should be no net exchange of energy between the photon and the pseudoscalar
gas. The rate at which photons loose energy by conversion to pseudoscalars
is given by
\begin{equation}
\dot\epsilon = {n_e n_\gamma\over \rho} {\Big<} c\int E_\phi {d\sigma\over 
dE_\phi} dE_\phi {\Big >} 
\label{eq:energyrate}
\end{equation}
where $\rho$ is the density of the medium and
the angular brackets denote thermal averaging. Equating this to the
rate at which photons are gaining energy due to the conversion of pseudoscalars
into photons we find
\begin{equation}
{n_e n_\gamma\over \rho} {\Big<} c\int E_\phi {d\sigma\over 
dE_\phi} dE_\phi {\Big>} 
={n_e n_\phi\over \rho} {\Big<} v\int E_\gamma {d\sigma\over 
dE_\gamma} dE_\gamma{\Big> }
\label{eq:balance}
\end{equation}
It is clear that the maximum temperature that the pseudoscalars
can acquire is equal to the photon temperature. The pseudoscalars would acquire
this temperature if in thermal equilibrium with the photon gas. At 
the temperatures corresponding to the solar interior 
the pseudoscalars would be relativistic and their 
number density would be equal to the photon number density.
This argument gives us a number
density in the core of the sun to be of the order of $10^{23}$ cm$^{-3}$. 
This is really a lower bound since, as discussed later, the temperature of 
pseudoscalars may be lower. In this case their number density has to be 
higher in order to maintain steady state with the photon gas. 

\begin{figure}[!ht]
\begin{center}
\rotatebox{0}{\scalebox{.6}{\includegraphics{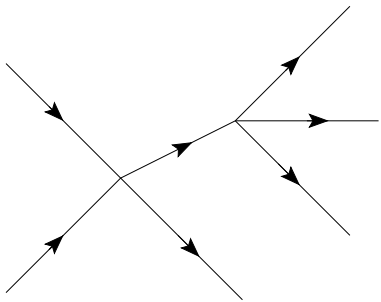}}}
\end{center}
\caption{ Feynman diagram for the fragmentation process 
$\phi \phi \rightarrow \phi \phi \phi\phi$.  }
\label{fig:feynmanncol}
\end{figure}

The pseudoscalars may not be in thermal equilibrium inside the sun. This
is due to the fragmentation process 
$$ \phi(k_1) + \phi(k_2) \rightarrow \phi(k_3) + \phi(k_4) + \phi(k_5) 
+ \phi(k_6) $$ 
which occurs through the Feynman diagram shown in Fig. 1. If we assume 
coupling $\lambda$ of order unity,
the cross section for this process at center of mass energy $E$, 
$\sigma\sim \alpha_\lambda^2 /E^2$, is much larger than  
the cross section for the inverse Primakoff process for the conversion of pseudoscalars
to photons. Hence pseudoscalars would fragment rather than convert to 
photons. The fragmentation process will stop only when the pseudoscalar
energy becomes comparable to their mass. Hence we expect the pseudoscalars 
inside the sun will have energies comparable to their mass and will
not be ultra-relativistic. 
 
The cross-section $\sigma$ for the process $\phi(k_1)+\phi(k_2)
\rightarrow \phi(k_3) + \phi(k_4)$ at leading order
in perturbation theory  is given by \cite{Jain06}
\begin{equation}
\sigma_{\phi \phi}=\frac{\alpha_{\lambda}}{8 E_{cm}^2}
\approx  1.2\times 10^{11}\alpha_\lambda \left[{10^{-6}\ {\rm
GeV}\over E_{cm}}\right]^2\ {\rm GeV}^{-2}
\label{eqn:cross}
\end{equation}
Assuming a center of mass energy of 1 KeV,
we find the mean free path $l \approx 2\times 10^{-7}/\alpha_\lambda$ cm. With  
$\alpha_\lambda=0.1$, this gives a mean free path $l\approx 2\times 10^{-6}$
cm, which is much smaller than the mean free path of photons inside the 
core. Hence we expect that pseudoscalars will contribute negligibly to
 radiative transport in comparison to photons and this mechanism
evades the limits imposed by considering cooling rate of the sun.

As seen above, the pseudoscalars are essentially trapped inside the sun and
unable to escape freely. Due to the fragmentation process they may not 
be in equilibrium with the rest of the system inside the sun. However
in steady state there would be no exchange of energy between pseudoscalars
and the remaining particles. Due to the fragmentation process a pseudoscalar
of energy $E$ will produce roughly $E/m_\phi$ pseudoscalars. The typical energy
of a pseudoscalar produced in the outer regions of the sun by conversion from 
photons is of order 1 eV. The temperature of the pseudoscalar gas is of order
of their mass, i.e. of order 0.001 eV. Hence a photon produced by pseudoscalars
will have energy of order 0.001 eV. Since the cross sections for Primakoff
and the inverse Primakoff process are roughly equal, we expect that, in steady
state, the number density of pseudoscalars in the outer regions of the sun
is of order 1000 times the photon density. The pseudoscalars will, therefore,
form a relatively cold and dense gas. For a certain range of self-coupling
it cannot be described as an ideal 
gas since its range of interaction may be much larger than the 
typical separation
between particles. 
The pseudoscalar gas will maintain steady state with the remaining particles
by producing large
number of low energy photons which will be quickly absorbed by the fermions 
inside the sun and hence will equilibrate.

Outside the sun the photons undergo free streaming. The pseudoscalars, 
however, are unable to propagate freely due to the background distribution of
pseudoscalars. We next get an estimate of the pseudoscalar density profile 
outside the sun by considering two limiting cases. 
We first assume that pseudoscalars
are in exact thermal equilibrium with the remaining particles inside the sun. 
In the other extreme we assume that the fragmentation process dominates. 

\subsection{Thermal Equilibrium}
Here we assume that pseudoscalars are in thermal equilibrium inside the
sun. This provides an important limiting case, although for the range of
couplings that we are studying this may never be realized. In this limit 
the pseudoscalars are ultra-relativistic near the surface of the sun. 
The number density of pseudoscalars is obtained by eq. \ref{eq:balance} and
is found to be same as that of photons throughout the radius of the sun.
We can compute the mean free path of pseudoscalars throughout the solar
radius using this density profile. It is found to be of order 
$10^{-7}/\alpha_\lambda$ 
cm near the core and increases to about $10^{-4}/\alpha_\lambda$ near
the solar surface. Hence for a wide range of values of $\alpha_\lambda$
it is much smaller than the photon mean free path and the energy transport
will occur primarily by pseudoscalar emission. 
Let $R_\odot$ denote the solar radius. For $r>R_\odot$, we assume,
for simplicity,
that we can describe pseudoscalars as an ideal Bose gas. At $r=R_\odot$
we impose the boundary condition that the pseudoscalar temperature is
equal to the temperature at the surface of the sun. 
The equation for the pressure gradient can be written as \cite{Weinberg}
\begin{equation}
{dP\over dr} = -{GM(r)\over r^2}(\rho_\phi(r) + P(r))\ ,
\label{Eq_Relativistic}
\end{equation}
where $\rho_\phi$ is the energy density of the pseudoscalars, 
$P$ the pressure and
\begin{equation}
M(r) = 4\pi \int_0^r dr'r'^2\rho_\phi(r')
\end{equation}
We point out that we are interested in the solution for $r>R_\odot$, where
we neglect all other particles except the pseudoscalars.
In eq. \ref{Eq_Relativistic} we have kept the dominant relativistic 
corrections. For $r>R_\odot$, we have $P(r) = \rho_\phi(r)/3$ and we may assume 
that $M(r) = M_\odot$. The solution can be written as
\begin{equation}
\rho_\phi(r) = \rho_\phi(R_\odot) \exp\left[-4GM_\odot\left({1\over R_\odot}
-{1\over r}\right)\right]
\end{equation}
This essentially means that $\rho_\phi(r)$ remains roughly constant inside
the solar system. The solution is not valid as we go to distances 
beyond the solar system since then we need to take into account gravitational
effects of other objects in the milky way. The pseudoscalar density profile
obtained for $r>R_\odot$ again implies a very small mean free path in this
region. Hence the pseudoscalars are unable to propagate freely throughout
the entire solar system and the standard solar model remains unaffected 
by the presence of these particles. 

\subsection{Fragmentation}
We next consider the other extreme where the fragmentation dominates. 
In this case we assume that the energy  per particle of the pseudoscalars  
is of the order of their mass due to the fragmentation process. 
The pseudoscalars may, therefore, not be ultra-relativistic. 
In order to get an estimate of the density profile for $r>R_\odot$ we 
assume that they are non-relativisitic in the outer regions of the sun, 
such that their temperature is smaller than their mass. 
The temperature of the photon gas near the surface is about $10^4$ K. Since
the mass of the pseudoscalar is about $10^{-3}$ eV, we expect that the 
pseudoscalar temperature $T_\phi < 10$ K. By using Eq. \ref{eq:balance}
and the fact that the cross sections for the direct and the inverse Primakoff
process are roughly equal, we find that the number density of pseudoscalars at
the surface $n_\phi(R_\odot) > 10^3n_\gamma$. The factor $10^3$ arises
due to the ratio $T_\gamma/T_\phi$. We may similarly obtain an estimate
of the density profile inside the entire radius of the sun. The precise value
is not relevant for our purpose. The number density is certainly atleast
as large as
the photon number density throughout the sun and that is enough to suppress
the energy loss through pseudoscalar emission to negligible values.   

\begin{figure}[!ht]
\begin{center}
\rotatebox{-90}{\scalebox{.5}{\includegraphics{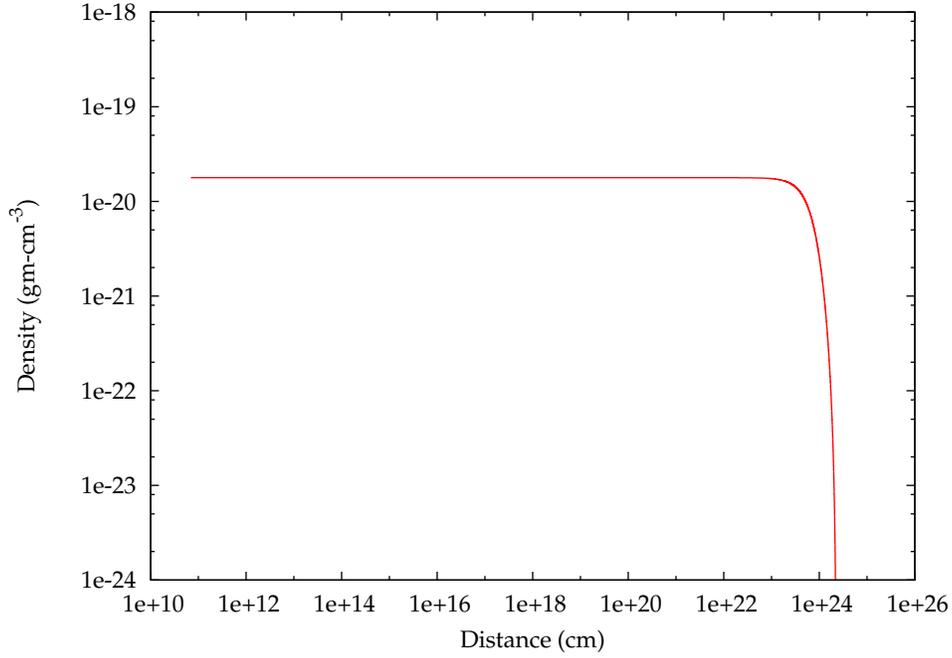}}}
\end{center}
\caption{  Density Profile of Pseudoscalars With Distance.  }
\label{fig:graph}
\end{figure}

The equation for hydrostatic equilibrium in this case is given by
\begin{eqnarray}
\frac{d P }{d r}= -\frac{GM(r)\rho_\phi(r)}{r^2}
\label{eq:one}.
\end{eqnarray}
We next need to specify the nature of change that the medium undergoes. We 
shall be interested in the pseudoscalar gas for $r>R_\odot$. In this region
we assume that it is governed entirely by its self-coupling and its interaction
with other particles is negligible. Due to the large cross section of
the process $\phi\phi\rightarrow \phi\phi$ it is clear that radiative 
transport of energy is negligible. This is because pseudoscalars
are not able to propagate freely in this medium. The medium is therefore
likely to undergo adiabatic changes and hence will reach a steady state 
such that $P\rho_\phi^{-\gamma}={\rm const}$, where $\gamma\equiv c_P/c_V$.  
For generality we assume a polytropic equation of state 
\begin{eqnarray}
P\rho_\phi^{-\gamma'}= \cal{K}
\label{eq:polytropic}
\end{eqnarray}
which models a large number of processes for different values of the index
$\gamma'$. For adiabatic change $\gamma'=\gamma$. We make the change of
variables $\rho_\phi(r)^{(\gamma' -1)}=z(r)$ and $x = r/\sqrt{\alpha}$ where
 $\alpha= {\cal{K}\gamma'}/[4\pi(\gamma'-1) G] $. In terms of $x$ and 
 $z$ the equation 
for hydrostatic equilibrium can be brought into the form of 
Lane-Emden equation,
\begin{equation}
\frac{1}{x^2} \frac{d}{dx} \left(x^2
\frac{dz(x)}{dx} \right)+[z(x)]^{{1}/{(\gamma' -1)}} = 0.
\label{eq:Lane-Emden}
\end{equation}
In order to bring the equation into dimensionless form it is convenient to 
use the scaled variables $\bar z$ and $\lambda$, defined by,
\begin{eqnarray}
z(x)&=&\bar{z}(x) \rho_\phi(R_\odot)^{(\gamma -1)} \cr
x&=&\lambda \rho_\phi(R_\odot)^{(\gamma -2)/2}
\label{s:1}
\end{eqnarray}
where $\rho_\phi(R_\odot)$ is the pseudoscalar density at the surface of the 
sun. The equation in terms of $\bar z$ and $\lambda$ is same as eq. 
\ref{eq:Lane-Emden} with $x$ and $z$ replaced by $\lambda$ and $\bar z$
respectively. The boundary conditions to be imposed are 
\begin{eqnarray}
\bar z(\lambda_o) &=& 1 \cr
{d\bar z\over d\lambda} (\lambda_o) &=& \delta
\end{eqnarray}
where $\lambda_o$ is the value of the dimensionless variable $\lambda$ at
the surface of the sun $r=R_\odot$ and $\delta$ can be determined by 
relating $d\bar z/d\lambda$ to $d\rho_\phi/dr$ at the surface of the sun. 
For the values given above, we find $\delta \approx 1\times 10^{-40}$.  
The resulting density profile is plotted in Fig: \ref{fig:graph}.
In obtaining this result we have set $\gamma'=\gamma=5/3$. In the present
case the medium is essentially opaque to the propagation of pseudoscalars
due to the large scattering cross section of the process $\phi\phi\rightarrow
\phi\phi$. Hence we can safely assume negligible energy transport and use
the adiabatic equation of state. 

We find that the density distribution is constant from the edge of the sun 
upto distance of order 100 Kpc. Ofcourse our solution is no longer valid once
we are outside the solar system since we have neglected the presence of
other objects in the milky way. Beyond a certain distance the pseudoscalar
density has to match the galactic pseudoscalar density. However we do not
address the issue of galactic density profile in this paper. 
We point out that even at very large distances from the sun, where the
pseudoscalar density reaches its galactic values, the mean free path of
pseudoscalars is still very small, of order of a cm, for a suitable choice of 
parameters $\alpha_\lambda$ and $\xi$. Hence even here pseudoscalars may not
be able to propagate freely for a wide range of these parameters. 

It is easy to see that the density
has to remain constant for a large range of values of 
$r$ above the surface of the
sun. For this purpose we start with the basic equation for pressure gradient, 
eq. \ref{eq:one}. Using the polytropic equation of state, 
eq. \ref{eq:polytropic}, we convert this into a differential 
equation for $\rho_\phi$
\begin{equation}
\frac{d \rho_\phi }{d r}= -\frac{GM(r)}{r^2{\cal K} \gamma}
(\rho_\phi(r))^{2-\gamma}
\end{equation}
The total mass contained within a distance $r$, $M(r)$, may 
be split up as
\begin{equation}
M(r) = M_\odot + \int_{R_\odot}^r 4\pi r'^2 dr' \rho_\phi(r')
\end{equation}
The second term on the right hand side is negligible compared first term 
as long as $r$ is comparable to or only a few orders of magnitude larger 
than $R_\odot$. We may, therefore, drop the second term for a wide range 
of values of $r$. With this approximation the equation can be integrated
analytically and we find
\begin{equation}
\rho_\phi(r)^{\gamma-1} - \rho_\phi(R_\odot)^{\gamma-1}
= - (\gamma-1){G M_\odot\over {\cal K} \gamma}\left({1\over R_\odot} - 
{1\over r}
\right)
\end{equation}
We find that the right hand side is negligible for $r>R_\odot$ and hence for a 
wide range of values of $r$ the solution is 
$\rho_\phi(r) \approx \rho_\phi(R_\odot)$.

The number densities obtained for all values of $r$ are again sufficiently large
so that the mean free path of pseudoscalars is much smaller than that of 
photons. Hence the energy loss is expected to arise primarily through photon
emission and the standard solar model will remain unaffected.  

It is interesting that the form of the solution for a wide range of values 
of $r$ is same as that obtained in section 2.1 
assuming thermal equilibrium throughout 
the sun. The number density is essentially constant within 
the entire solar system for $r>R_\odot$. The precise number is obtained by the 
boundary condition
imposed at the surface of the sun. Furthermore our mechanism for evading
astrophysical bounds on pseudoscalar coupling and mass values is found
to be applicable independent of the assumptions used to obtain the density
profile inside the sun. 

We next check if our solution is consistent with the constraints imposed
by the Pioneer anomaly \cite{Pioneer}. We find that the solution obtained
is not in conflict with the maximum allowed dark matter density inside 
the solar system \cite{Slava,Foot}. However the density is about an order of 
magnitude small compared to what is required to reproduce the anomalous 
acceleration of pioneer \cite{Pioneer} crafts 10 \& 11, towards Sun. We 
may achieve higher densities if the pseudoscalars have 
temperature of the order of $1/2$ K at the surface of the sun, instead of
10 K, which we assumed. This can arise if the pseudoscalars also loose energy
as the propagate inside the sun \cite{Jain06}. By using Eq. \ref{eq:balance}
we find that this gives a higher pseudoscalar density at the surface of the
sun by a factor of about 20. This leads to an increase in 
the density in such a way that it exactly matches the pioneer anomaly data.

\section{Medium modifications}
We next address an important consistency check of our proposal. It is important
to determine how the presence of background pseudoscalars affect the results
of the PVLAS experiment. The density of pseudoscalars is relatively small
and due to their weak coupling with photons, they will not significantly
affect the propagation of photons. The dominant effect is modification
of the pseudoscalar self energy. The 
medium modifications can be taken into account by using the modified mass
$m_T^2 = m_\phi^2 + \delta m_{\rm th}^2$, where the thermal mass \cite{Das},
\begin{equation}
\delta\rm{m}^2_{\rm{th}}=\frac{\lambda}{4\pi^2}
\int^{\infty}_{0}k^2 dk  \frac{1}{ E_{k}}
\frac{1}{e^{\beta E_k} - 1} .
\label{M2}
\end{equation}
where $E_{k}=\sqrt{k^2+m^2_{\phi}}$ and $\beta = 1/T$. 
In the high temperature limit, i.e., $T>>m_{0}$, the result is straightforward 
and it turns out to be
\begin{eqnarray}
\delta\rm{m}^2_{\rm{th}}=\frac{\lambda T^2}{24}
\label{M3}
\end{eqnarray}
In the low temperature limit $\beta\rightarrow \infty$, we find
\begin{equation}
\delta\rm{m}^2_{\rm{th}} \sim \frac{\lambda}{4\pi^2}
\sqrt{\pi mT\over 2}
\,Te^{-m/T}
\label{low-temp-limit}
\end{equation}

At the position of earth, we expect that the
temperature would be at most as high as the mass of the pseudoscalar particle,
due to the fragmentation process.
Hence the change in the pseudoscalar mass due to thermal effects 
would be small as long as $\lambda<1$.  
The modification in mass could be large if for some
reason the fragmentation process is not very effective. In this case we should
interpret the mass extracted by the PVLAS experiment as being dominantly the
thermal mass. In either case our mechanism for evading astrophysical 
bounds remains applicable.

\section{Summary and Conclusions}
We have shown that the solar astrophysical limits can be evaded provided we
assume a sufficiently large value of the self-coupling of 
pseudoscalars and that pseudoscalars form some component of the 
galactic dark matter. The self-coupling
required is within the perturbative regime and the galactic density of
pseudoscalars is assumed to be a negligible fraction of the dark matter density.
Inside the sun the 
maximum possible value of pseudoscalar temperature at any radius $r$ is 
equal to the
solar temperature at that point. The corresponding mean free path of 
pseudoscalars is found to be sufficiently small so that they 
contribute negligibly
to radiative transfer. 
We have shown that outside the sun the density profile
of pseudoscalars is roughly uniform. 
The density outside the star is found to be such that the mean free
path of pseudoscalars is smaller than a cm for a wide range of values of
the self coupling $\alpha_\lambda$. 
Hence the pseudoscalars cannot escape freely from the sun
and the standard solar model remains unaffected. 
The solar limits on the pseudoscalar-photon coupling are, therefore, evaded.

\end{document}